\documentstyle[11pt,aaspp4,epsfig]{article}
\def\ls{{_<\atop^{\sim}}}
\def\gs{ { _>\atop^{\sim} } }
\def\cgs{ ${\rm erg~cm}^{-2}~{\rm s}^{-1}$ } 

\begin{document}
\title{The BeppoSAX High Energy Large Area Survey (HELLAS)}
\author{F. Fiore$^{1,2,3}$, P. Giommi$^1$, 
F. La Franca$^4$, G. Matt$^4$, G.C. Perola$^4$, 
A. Comastri$^5$, S. Molendi$^6$, M.~Elvis$^3$, I Gioia$^{7,8}$
F. Tamburelli$^1$, D. Ricci$^1$, F. Pompilio$^4$}
\affil{
$^1$ BeppoSAX Science Data Center, via Corcolle 19, I-00131 Roma, Italy \\
$^2$ Osservatorio Astronomico di Roma, via Frascati 33, I-00040 Monteporzio, 
Italy \\
$^3$ Smithsonian Astrophysical Observatory, 60 Garden Street, 
Cambridge, MA 02138\\
$^4$ Universita' Roma Tre, Via della Vasca Navale 84, 
I-00146 Roma, Italy \\ 
$^5$ Osservatorio Astronomico di Bologna, via Ranzani 1, I-40127
Bologna, Italy \\
$^6$ IFCTR/CNR, via Bassini 15, I-20133 Milano Italy\\
$^7$ Institute for Astronomy, Hawaii, 96822 USA
$^8$ IRA/CNR, Bologna, Italy
}

\begin{abstract}
We have surveyed $\sim 50$ deg$^2$ of sky in the largely unexplored
5-10 keV band using the BeppoSAX MECS instrument, finding 180 sources.
After correction for the non uniform sky coverage we find about 18
sources deg$^{-2}$ with $F_{5-10keV}\gs5\times10^{-14}$ \cgs, and
resolve 30-40 \% of the hard 5-10 keV Cosmic X-ray Background (XRB).
Optical identification of a first small sample of sources show that
most (11 out of 14) are AGN.  Six of these show evidence of
absorption/extinction in X-ray/optical, thus providing support to the
scenario in which the hard XRB is largely made by obscured AGN (Setti
\& Woltjer 1989, Comastri et al.  1995).
\end{abstract}

\section{Introduction}

Hard X-ray selection is an efficient way to separate sources
accretion-powered, such as Active Galactic Nuclei (AGN), from sources
dominated by starlight emission, which heavily populate optical and IR
surveys.  Hard X-rays are also less affected by selection biases than
other bands. For example, optical and UV color selection is often
biased against objects with even modest extinction, or with an
intrinsically `red' emission spectrum.  Soft X-ray selection is biased
against highly obscured objects, a column density of a few times
$10^{22}$ cm$^{-2}$ is sufficient to reduce by $\sim100$ times nuclear
emission below 2 keV. The same column has negligible effect in the
5-10 keV band.  This band is now accessible to surveys, thanks to the
imaging instruments on board the ASCA and BeppoSAX satellites.

Hard X-ray selection permits us to: a) directly resolve the sources
making most of the hard XRB (the energy density of the XRB at 5-10 keV
is 2-3 times higher than at 1 keV, where previous X-ray surveys have
provided most information).  b) Measure the fraction of obscured to
unobscured objects or, more generally, the distribution of absorbing
columns of AGN.  c) Search for previously `rare' AGN, like `red'
quasars or other minority AGN populations (Kim \& Elvis 1998) and
quantify their fractional contribution to the AGN family.  This
allows us to start tackling the following open questions: (1) Is the
hard XRB mostly made by AGN, like the soft XRB?  (2) Is the XRB
characteristic shape due to a large population of absorbed AGN? (3)
Can we distinguish between different scenarios?  AGN synthesis models
for the XRB based on unification schemes (Antonucci 1993) assume the
same evolution for type 1 and type 2 objects and predict a sizeble
population of `quasar~2' (with $L_X>10^{44}$ erg s$^{-1}$ and
log$N_H>$23).  Alternatively, the bulk of the hard XRB could be made
by a large population of lower luminosity Seyfert 1.8-2 like galaxies,
as in scenarios where the absorption takes place in a starburst region
surrounding the nucleus (Fabian et al. 1998). In this case the amount
of the absorbing gas may depend on the nuclear mass and consequently also
on its luminosity.  The evolution of the type 2 AGN luminosity function would
then be strongly affected by the nuclear environment and may imply a
link between AGN evolution and the history of star-formation (Madau et
al. 1996).

\section {The HELLAS survey}

We decided to perform a high energy survey using the BeppoSAX (Boella
et al. 1997a) MECS X-ray imaging instrument (Boella et
al. 1997b). Details can be found in Fiore et al. (1998a,1999).  The
HELLAS survey has been performed in the hard 5-10 keV band for the
following reasons. 1) This band is the closest to the peak of the XRB
energy density that is reachable with the current imaging X-ray
telescopes. 2) The BeppoSAX MECS PSF greatly improves with energy: in
the 5-10 keV band it is a factor of $\sim2$ sharper than in the softer
1-5 keV band (providing a 95 \% error radius of 1 arcmin), which allows
easier optical identification of the sources. 3) Including the softer
1-5 MECS range only increases the background for heavily cut-off
sources with few photons below 5 keV, thus reducing the chances of
detecting faint hard sources.  About 180 sources have been detected at
a confidence level $\gs 3.5\sigma$.  The count rates were converted to
fluxes using the HR=4.5-10 keV/1.3-4.5 keV hardness ratio to estimate
the spectral shape.  A logN-logS function has been computed excluding
from the sample the sources detected at offaxis angles $>22$ arcmin or
near the strongback support of the MECS beryllium window, to minimize
possible systematic errors due uncertainties in the MECS calibration
in these regions.  Figure 1 shows the cumulative 5-10 keV logN-logS
after correction for the energy dependent sky coverage.  We find about
18 sources deg$^{-2}$ at our flux limit of $F_{5-10~keV} =
5\times10^{-14}$ \cgs (also see Giommi et al 1998).  This logN-logS
corresponds to a resolved fraction of the 5-10 XRB equal to 30--40 \%,
depending on the XRB normalization (see Comastri 1998).  18 sources
deg$^{-2}$ is similar to the space density of optically selected
quasars at B=20-21 (Zitelli et al 1992).  AXAF and XMM will reach
fluxes 500--100 times fainter, respectively, than BeppoSAX and ASCA,
and so will find a density of sources higher or comparable to that of
AGN surveys performed in any other band, even if the X-ray logN-logS
flattens below $5\times 10^{-14}$ to a slope between 0.5 and 1 (as
suggested by fluctuation analysis of ROSAT HRI deep fields, Hasinger
et al. 1998).

We have divided the HELLAS sample into three groups of sources,
depending on their spectral hardness. The three resulting logN-logS
are shown in Figure 1, along with the prediction for the 5-10 keV band
obtained by extrapolating the ROSAT 0.5-2 keV logN-logS assuming an
unabsorbed power law spectrum of $\alpha=1$. This extrapolation falls
short of the total observed logN-logS by a factor of $\sim5$ but is
close to the logN-logS of the softer HELLAS sources, in agreement with
the suggestion that the hard X-ray source population is dominated by
obscured sources.
 
\section{Optical spectroscopic identifications}

Ten percent of the HELLAS sources have been identified with
previously know AGN (13), clusters of galaxies (3), stars (1) and
Cataclismic Variables (1).  The 13 AGN were discovered in radio, soft
X-ray, and optical surveys, and therefore this sample is biased against
highly absorbed sources. To remedy this bias we have started a program
to spectroscopically identify the rest of the HELLAS sample.
Between 1 and 7 optical candidates were identified on APM scans of the
E (R) POSS plates down to R$\sim 20$m, within a conservative error-box
of 60$''$ radius around each X-ray source.  {\it Unlike previous
identification campaigns of ASCA sources, which take advantage of
correlations with ROSAT source catalogs or Radio catalogs, our
strategy was to obtain spectra of all candidates down to the chosen
R=20 magnitude limit. To avoid any possible selection bias we do not 
require the HELLAS sources to be detected also at other (softer) energies.}
We carried out spectroscopic identification of optical candidates for
the HELLAS sources using the RC spectrograph (RCSP) at the Kitt Peak
4m telescope for 9 sources, the FAST spectrograph at the Whipple 60''
telescope for 1 source, the Hawaii 88'' for 2 sources and
EFOSC2 at the ESO 3.6m for 2 sources (Fiore et al. 1998b).

%\begin{figure}[h]
%  \epsscale{1.0}
%  \begin{center}
%%    \plotfiddle{epic_qe.eps}{250pt}{270}{80}{80}{-300}{430}
%%   \plotfiddle{epsfile}{vsize(pt)}{rot(deg)}{hsf(%)}{vsf(%)}{htrans(pt)}{vtrans(pt)}
%    \epsfig{figure=fiore_lnls5-10.ps, height=10.cm, width=14cm,angle=0}
%  \end{center}
%  \caption{Figure 1:
%HELLAS 5-10 keV total logN-logS (thick line).
%The dotted, dashed and dot dashed lines shows the logN-logS 
%of the hard (HR$<-0.27$, roughly corresponding to a column of 
%log$N_H>$23 for a power law spectrum of $\alpha_E=0.8$ and z=0),
%intermediate (-0.27$<$HR$<$0.27), and soft 
%(HR$>0.27$, log$N_H<$22) HELLAS sources.
%The solid thin line shows the extrapolation from the ROSAT 0.5-10 keV
%logN-logS assuming an unabsorbed power law spectrum with $\alpha_E=1$.
%}
%
%\end{figure}

\begin{figure}[h]
  \epsscale{1.0}
  \begin{center}
    \epsfig{figure=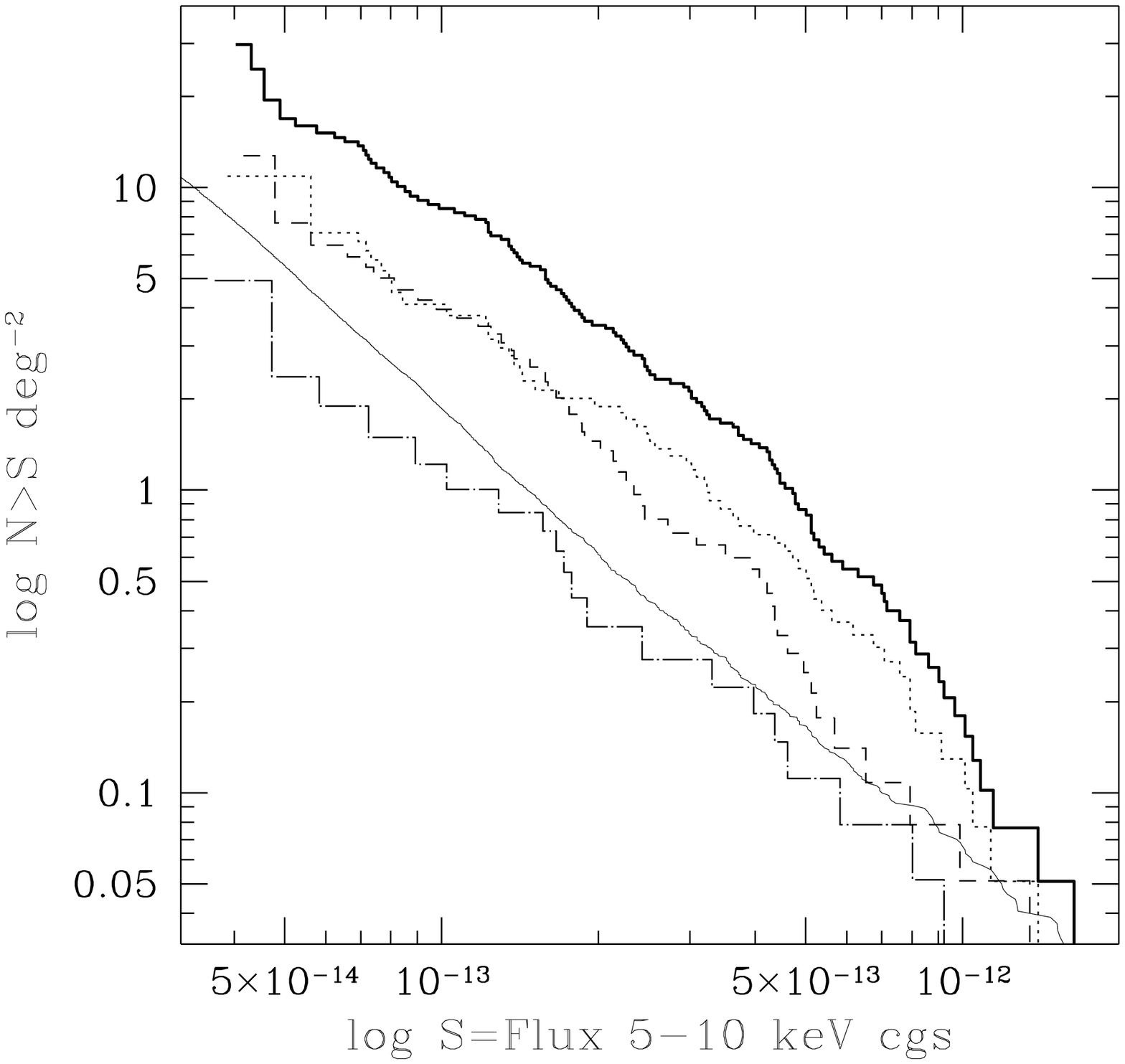, height=7.cm, width=8cm,angle=0}
    \epsfig{figure=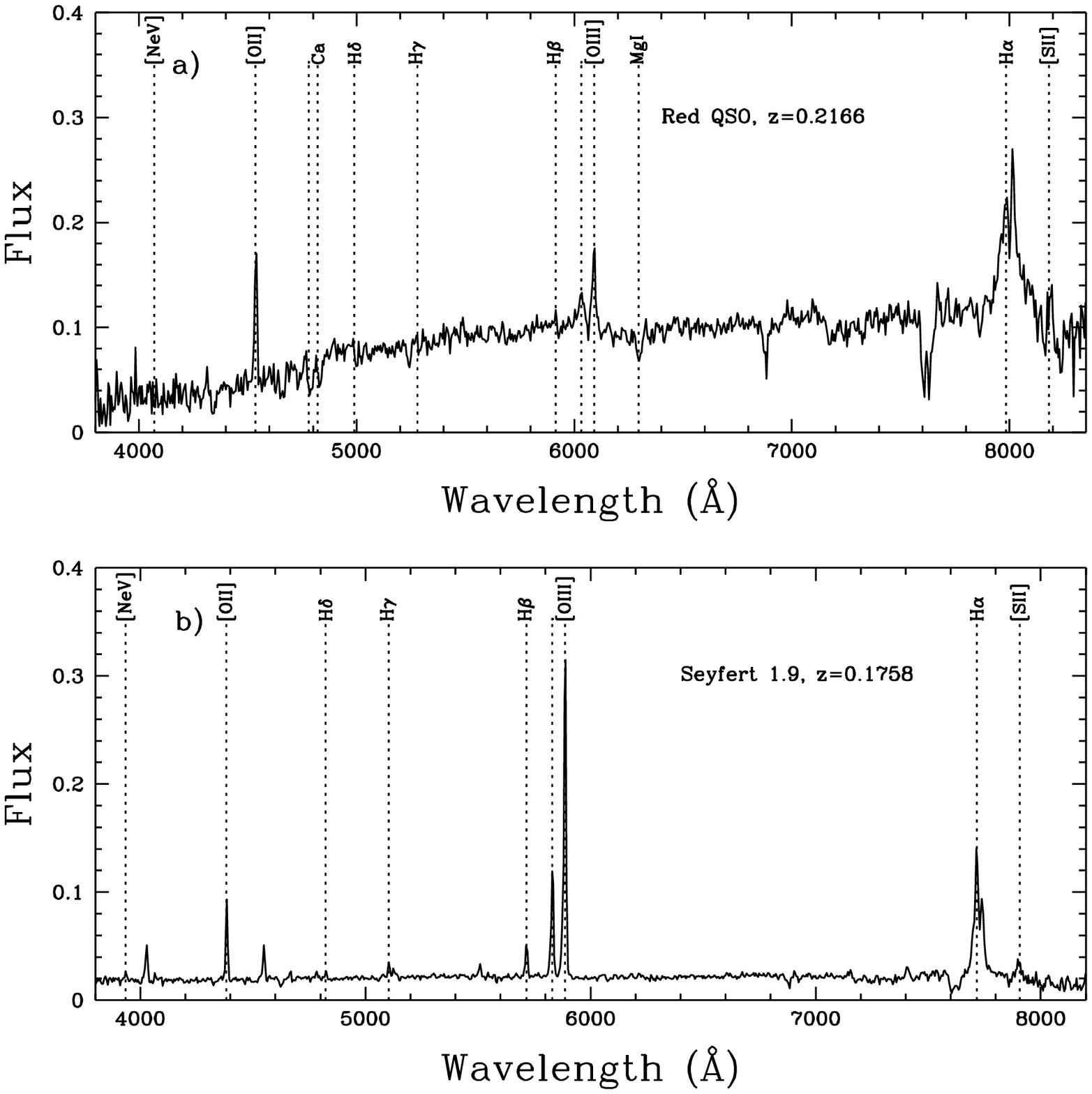, height=9.cm, width=8cm,angle=0 }
  \end{center}
  \caption{
{\bf Figure~1}:
HELLAS 5-10 keV total logN-logS (thick line).
The dotted, dashed and dot dashed lines shows the logN-logS 
of the hard (HR$<-0.27$, roughly corresponding to a column of 
log$N_H>$23 for a power law spectrum of $\alpha_E=0.8$ and z=0),
intermediate (-0.27$<$HR$<$0.27), and soft 
(HR$>0.27$, log$N_H<$22) HELLAS sources.
The solid thin line shows the prediction from the ROSAT 0.5-2 keV
logN-logS, assuming an unabsorbed power law spectrum with $\alpha_E=1$.
{\bf Figure~2}:
The spectra of the type 1.9 AGN 1SAXJ1218.9+2958 and 
of the red quasar 1SAXJ1353.9+1820 
}
\end{figure}

The observations of the fourteen error-boxes led to 13 new
identifications (Fiore et al. 1998b).  Five AGN are normal broad line
``blue'' quasars ($0.2<$z$<1.3$). Two AGN are broad line `red' quasars
($\alpha\sim-3$, B-R=2.9 and 2.6, $0.2<$z$<0.4$).  Four objects at
$0.04<$z$<0.34$ have an [OIII]/H$_{\beta}$ flux ratio of about 10,
typical of Seyfert 2 galaxies, but show broad ($\sim 4000$ km
$s^{-1}$) H$_{\alpha}$ or H$_{\beta}$ wings.  We identified these
objects as intermediate type 1.8-1.9 AGN. Two narrow emission line
galaxies ($0.08<$z$<0.18$) have diagnostic line ratios indicating
lower excitation than in Seyfert 1.8-2 galaxies but higher than in HII
galaxies.  We classify them as LINERS.  One of the fourteen HELLAS
sources observed still remain unidentified.  Figure 2 shows the
spectra of a `red' quasar and of a type 1.9 AGN. These first
identifications confirm that most of the HELLAS sources are indeed
AGN. The source breakdown is however rather peculiar.  In addition to
five normal broad line quasars, typically found in optical and soft
X-ray surveys, we found six ``intermediate'' AGN and two LINERS.
These objects are ``intermediate'' both for their optical spectra
which places them between type 1 and type 2 AGN and because of their
X-ray absorption, which is typically log$N_H$=22.5-23 (estimated from
hardness ratios).  `Red' quasars have been selected in the past in the
radio band (Smith \& Spinrad 1998, Webster et al. 1995) or in soft
X-rays (provided that the absorbing column is not too thick,
A$_V\ls2$, log$N_H<21.7$, Kim \& Elvis 1998). Their number density
relative to normal ``blue'' quasars is thought to be no more than a
few percent. In contrast we found a fraction of about 20 \%.  Hard
X-rays selection appears to be an efficient way of discovering `red'
quasars over a broad range of column densities.  A comparison with
ROSAT and ASCA is telling: the fraction of AGN showing evidence of
obscuration of the nucleus in either optical or X-rays (6 out of 11
high excitation AGN) is $0.55^{+0.11}_{-0.14}$ (Gehrels 1986, 67 \%
confidence intervals).  Including the two LINERS brings this fraction
to $0.62^{+0.10}_{-0.11}$.  The fraction of intermediate AGN + narrow
line galaxies in the ROSAT 0.5-2 keV survey of the Lockman hole
(Schmidt et al. 1998) is 0.26$\pm$0.09 (10 out of 39 objects).  The
same fraction in the ASCA-ROSAT combined surveys of Boyle et al.
(1998) and Akiyama et al. (1998) is 0.33$\pm$0.08 (20 out of 61
objetcs).  The HELLAS fraction including (excluding) LINERS is
different from the ROSAT and ASCA-ROSAT fractions at the 99 \% and 95
\% (92 \% and 80 \%) confidence level respectively.  These differences
are probably due the fact that ROSAT and ASCA-ROSAT surveys, using
soft X-ray data, are biased against faint highly obscured objects.

\section{Conclusions}

Hard X-ray surveys are the most efficient and unbiased way to search
for any form of accretion process in the Universe.  The advent of
imaging instrument sensitive above 2 keV opens up new frontiers in the
study of the X-ray sky, in that they allow us to go ``deeper''
(i.e. to find sources $\sim 100-1000$ fainter than in the HEAO-1,
EXOSAT and {\it Ginga} surveys) and explore nearly untouched
``planes'' in discovery space (e.g. absorption in sources at z$>0.1$,
extreme spectral shapes).  We have started a program to identify a few
hundred hard (5-10 keV) X-ray selected sources.  Preliminary results
based on a small sample of sources show that most (11 out of 14) are
AGN, half of which showing evidence of absorption at the level
predicted by AGN synthesis models of the XRB (Comastri et
al. 1995). The first identifications are also providing surprises.  In
addition to normal broad line blue quasars we are finding a few
LINERS, `red' quasars, and several ``intermediate'', relatively low
luminosity ($L_X=10^{43}$ erg s$^{-1}$, M$_B$=-20) AGN.  This rich
diversity contrasts with the uniformity of spectra found in optical/UV
and soft X-ray surveys, and suggest that the AGN selection criteria
inherited from these works have restricted our view of the true range
of AGN spectra (Elvis 1992).  The AGN phenomenon, e.g. accretion, is
likely to manifest itself in a wider and more complex way than we used
to know.

We did not find any luminous, narrow line and highly absorbed quasar
(the so called `quasar~2').  This is not surprising, since their
number, as predicted by the Comastri et al. (1995) XRB model at our
flux limit, is 0.4 deg$^{-2}$, i.e only 5 among the 180 HELLAS
sources. However, other recent studies (Boyle et al. 1998, Akiyama et
al., these proceedings) also failed to find objects of this kind.
Although the statistics are not yet good enough to reach strong
conclusions, these results do suggest that the bulk of the hard XRB
could be made by a large population of lower luminosity Seyfert 1.8-2
like galaxies and moderately absorbed `red' quasars, i.e. `minority'
objects in optical and soft X-ray surveys. In this regard, it is worth
remarking that a sizeble population of low luminosity AGN (M$_B$ in
the range --17; --20) may exist at very faint optical fluxes (V=26-27)
in the Hubble Deep Field North (Jarvis \&MacAlpine 1998).  This
conclusion will be soon tested by the deep and high spatial resolution
surveys that will be performed by AXAF and XMM in the next 1--2
years. Deep observations of the Hubble Deep Fields with these
observatories should be able to find the X-ray emission from these
putative Active Nuclei.

\acknowledgements
We thank the BeppoSAX SDC, SOC and OCC team for the successful
operation of the satellite and preliminary data reduction and
screaning.  We thank in particular Alessio Matteuzzi for his work on
MECS source position reconstruction,  
Phil Massey for his help at the Kitt Peak 4m, and Perry Berlind 
for the Whipple 60'' spectra of the 1SAXJ1519.5+6535 field.

\end{document}